\documentclass{proc}
\usepackage{times}

\usepackage{amssymb}
\usepackage[hang,scriptsize]{subfigure}
\usepackage{latexsym}

\newcommand{\bvec}[1]{\mbox{\boldmath $\rm#1$} }

\usepackage{boxedminipage}

\usepackage[dvips]{graphicx}
 
\makeatletter
\def\@maketitle{%
  \vbox to 2.3in{%
    \hsize\textwidth
    \linewidth\hsize
    \vspace*{1.5cm}
    \centering
    {\bfseries\huge \@title \par}
    \vskip 2em
    {\large \begin{tabular}[t]{c}\@author \end{tabular}\par}
    \vfill}    \vspace*{1.0cm}
}
\renewcommand\section{\@startsection {section}{1}{\z@}%
     {.7\baselineskip plus\baselineskip}{.5\baselineskip}
                                   {\normalfont\Large\bfseries}}
\renewcommand\section{\@startsection {section}{1}{\z@}%
      {.5\baselineskip\@plus.7\baselineskip}{.3\baselineskip}%
                                   {\normalfont\Large\bfseries}}
\renewcommand\subsection{\@startsection{subsection}{2}{\z@}%
       {.5\baselineskip\@plus.7\baselineskip}{.3\baselineskip}%
                                   {\normalfont\large\bfseries}}
\renewcommand\subsubsection{\@startsection{subsubsection}{3}{\z@}%
      {.5\baselineskip\@plus.7\baselineskip}{.3\baselineskip}%
                                     {\normalfont\normalsize\bfseries}}
\makeatother
 
\renewenvironment{abstract}%
  {\normalfont
    \list{}{\labelwidth0pt
      \leftmargin0pt \rightmargin\leftmargin
      \listparindent\parindent \itemindent0pt
      \parsep0pt
      
    }%
    \item[\hskip\labelsep\bfseries\abstractname\enspace --] \itshape%
}{%
  \endlist}

\newcommand{\keywordsname}{Keywords}
\newenvironment{keywords}%
  {\normalfont
    \list{}{\labelwidth0pt
      \leftmargin0pt \rightmargin\leftmargin
      \listparindent\parindent \itemindent0pt
      \parsep0pt
      }%
    \item[\hskip\labelsep\bfseries\keywordsname:]}{\endlist}

\newcommand{\bi}{\begin{itemize}}
\newcommand{\ei}{\end{itemize}}
\newcommand{\be}{\begin{equation}}
\newcommand{\ee}{\end{equation}}
\newcommand{\bea}{\begin{eqnarray}}
\newcommand{\eea}{\end{eqnarray}}
\newcommand{\ba}{\begin{array}}
\newcommand{\ea}{\end{array}}

\begin{document}
\title{Determining possible avenues of approach using ANTS}
\author{
Pontus Svenson ~~~~~~~ Hedvig Sidenbladh\\[0.3cm]
Department of Data and Information Fusion\\
Division of Command and Control Systems\\
Swedish Defence Research Agency\\
SE 172 90 Stockholm, Sweden\\
{ \tt ponsve,hedvig@foi.se}\\
{ \tt http://www.foi.se/fusion/}
}
\maketitle
\begin{abstract}

Threat assessment is an important part of level 3 data fusion.
Here we study a subproblem of this, worst-case risk assessment.
Inspired by agent-based models used for simulation
of trail formation for urban planning,
we use {\em ant 
colony optimization (ANTS)} to determine possible
avenues of approach for the enemy, given a situation picture.

One way of determining such avenues would be to calculate
the ``potential field'' caused by placing sources at possible
goals for the enemy. This requires postulating a functional
form for the potential, and also takes long time.
Here we instead seek a method for quickly obtaining an
effective potential.
ANTS, which has previously been used to obtain approximate solutions to
various optimization
problems, is well suited for this. The output of
our method describes possible avenues
of approach for
the enemy, i.e, areas where we should be prepared for attack.
(The algorithm can also be run ``reversed'' to instead
get areas of opportunity for our forces to exploit.)

Using real geographical data, we found that our method gives a fast
and reliable way of determining such avenues. Our method 
can be used in a computer-based command and control system to replace the 
first step of human intelligence analysis.
\end{abstract}

\begin{keywords}
{
ANTS, ant colony optimization, heuristical methods, 
swarm intelligence, determining avenues of approach, threat analysis, 
threat assessment, worst-case risk assessment}
\end{keywords}

\section{Introduction}
\label{intro}

Threat assessment is an important part of level 3 data fusion, as defined
in~\cite{handbook}. The goal of level 2 data fusion~\cite{handbook},
is to provide an accurate picture of current enemy activity. Given this 
information about the current time instant, the goal 
of threat assessment is to extrapolate it into the future to see which own 
objects are most threatened, and to determine which enemy objects pose the 
greatest threat. This is a complex task involving modeling 
the enemy's units, objectives, estimated knowledge about our
forces, and doctrines. The role of threat 
assessment in level 3 data fusion and its relation
to the worst-case risk assessment studied here is discussed 
further in section~\ref{threat}. 

The algorithm presented in this paper is designed to give a worst-case scenario
of what important locations the enemy objects can reach within certain time 
limits, given the terrain and the estimated mobility capabilities of the 
enemy objects. In section~\ref{application} we describe how the algorithm works. 

Many important optimization problems can not be solved exactly
in an efficient way (see, e.g.,~\cite{papadimitriou,gareyjohnson}).
For many problems there are, however, fast, approximate methods that
will in most cases give a solution that is ``good enough''.
One such algorithm,
building on ideas from biology, is 
{\em ant colony optimization (ANTS)}~\cite{dorigo,bonabeau}. In nature, ants 
communicate by deploying pheromone (smell) paths that indicate the way 
between the ant-hill and food supplies. In the ANTS algorithm, ants are 
simulated agents who move through a search space or energy landscape
looking for good locations. When an ant has found a sufficiently good 
place, it stops and distributes a pheromone along the path it took from its 
starting position. In our case, ants start from locations
of enemy units as given by level 2 data fusion, and their goals
are positions of own units. 
The ants will use the smell in addition to 
geographical and military-value information to determine
where to move. The ANTS algorithm is described in more detail in 
sections~\ref{ants} and~\ref{application}.
Since we assume that the enemy ants can determine
when they have reached an own force-location, the output of our method will be
a worst-case result.

Results of the ANTS method for a test scenario
are shown in sections~\ref{gottrora} and~\ref{presentation},
where we give some suggestions for how to present the avenues
of approach to a user. Section~\ref{potentialsektion}
compares the effective potential determined by ANTS with
a calculated potential.
Finally, sections~\ref{discuss}
to~\ref{conclusions} discusses our results and presents
some ideas for future extensions of the method.

\section{Background}
\subsection{Threat versus worst-case risk}
\label{threat}

Threat assessment is one of the most important and challenging
parts of an information fusion system. 
The method presented in this paper gives a very fast way of
obtaining possible avenues of approach
given a situation assessment.
It does not provide a complete
threat analysis. Rather, the output of our algorithm
is a map showing an ``intelligent guess'' as to what
areas of the map the enemy may reach. The method will
very quickly display a first guess and will then refine this
incrementally until it converges to an approximation of
the enemy's avenues of approach.
Our algorithm basically simulates many imaginary enemy units
doing local searches for own units;
this is what makes it a worst-case risk assessment tool.
An ``exact'' threat assessment would assume that the
enemy has uncertain information, assign relevant probabilities
or beliefs, and produce a ranked list of the enemy's
probable objectives and the paths to them. Such a system could use
our ant-based algorithm as a subsystem, combining the output of several
runs of this subsystem using, e.g., some sort of 
Dempster-Shafer~\cite{shafer} or
random set~\cite{randomsetbok} formalism. Given a 
Dempster-Shafer belief function
describing a situation picture, the ANTS method could be run
for each of its focal elements. Attaching the corresponding
probability mass to the resulting avenues would then produce a
belief function over avenues of approach.

Despite the heuristical nature of ANTS,
we argue that this kind of method is an important
and indeed necessary part of an operational information fusion
system.
Our ant-based risk assessor provides a real-time indication
of where enemy units might pose the biggest threat. 
Obtaining a threat analysis that is
provably correct or correct with probability $1-\epsilon$
using current methods requires a very large amount of computer
resources. Our method, in contrast, runs in real-time
and incrementally improves its output.
Even in a command and control system that has
a complete threat assessment module, an
approximate solution such as ours has its place.
The output of the fast, approximate algorithm can be used to aid
in quick decisions. By comparing the output of this algorithm with
the one from a slower but more accurate method, we can find
out if the first suggestion was indeed correct. This is very similar
to how humans often make a first guess, perhaps without knowing
all the facts, and later refine their answer after thinking about
the problem. It it also possible that a comparison between
the worst-case analysis of ANTS and the action actually taken 
by the enemy could help us in determining how much the
opposite side actually knows.

By having several different subsystems performing the same 
or nearly the same function,
we also gain robustness for the overall system --- some parts of
it can break down and it will still be able to function.
An additional benefit of having two or more subsystems performing the
same task is that if the different methods give different results, something
extraordinary (e.g., a completely new
military strategy is adopted by the enemy) might be taking place and 
human operators/analysts need to look at the
situation in more detail.

Our method can be used for worst-case risk assessment on all levels
of force aggregation.
It is
just as easy to determine where a single tank might be headed as it
is to determine where a battalion is going. Note that the 
geographical/locational information might actually be more accurate
for a battalion since there are more restrictions where it can go 
(e.g., it needs a corridor of certain width).

\subsection{Ants and Swarm Intelligence}
\label{ants}

ANTS is a form of collective, intelligent agent system. 
It is similar to other models for swarm intelligence
and crowd behavior (e.g.,~\cite{helbing,bonabeau:sciam}).
It was first used by Dorigo (e.g.,~\cite{dorigo}) to solve 
the traveling salesperson problem and has later been used for solving
problems ranging from
graph coloring~\cite{costa} to routing and
load balancing~\cite{schoonderwoerd,kassabalidis}.
The method shares a number of conceptual features
with models for describing complex behavior such
as Braitenberg vehicles~\cite{braitenberg} or Langton's 
vant~\cite{langton}.

ANTS is very similar to random walk or diffusion based
methods for solving optimization problems, but adds
interaction between the walkers to produce results
more quickly. An ant is
an agent that moves in the space of all solutions
to the problem. The movement is basically a random
walk but with one addition. An ant that reaches a good
solution to the problem distributes a smell along the path
it took to reach this solution. This smell is then sensed
by other ants and increases their probability to move
to a site with a high smell. This will lead to
many ants being attracted to good sites. This process
is similar to that used by real ants to communicate
paths to food-sources; hence the name ``ants'' for
the agents. ANTS is suited
primarily for problems that have a natural representation
in a low-dimensional metric space or a graph of not too high
connectivity.

Note that the way we use ANTS here is
different from traditional optimization.
We do not solve an optimization problem.
The smell, which gives the ants their ability
to solve optimization problems, is here the output of
the method. This smell determines avenues of approach,
that is, locations where the enemy
may move.

It is instructive to compare the ANTS method with tracking methods.
The model for how the agents move is similar to that
used in for example particle filtering (e.g.,~\cite{gordon93}), 
but adds some extra
non-linearity in the form of the use of the smell. This
leads to a different, more
explicit, kind of interaction between the agents than
in tracking models. In a way, we can say that the smell
takes the place of observations when we attempt to predict future
positions.

Active walker models similar to the ants used here have previously
been used to model formation of trails and crowd
behavior (e.g.,~\cite{helbing}). One application of this which
is somewhat similar to our risk analysis is in urban 
planning. Another possible application of methods such as these is
to crowd-control; here ANTS would be used to predict
where a crowd will move.

\section{The ANTS algorithm}
\label{application}

Our method is best explained by examining a situation with
one enemy unit. Start by inserting $N$ ants
at its location. In each time-step, each ant randomly selects
a neighbor of its current location and moves there. The probability
is not uniform over all neighbors. Instead, the type of the
unit represented by the ant and the terrain is taken into account
so that, e.g., it is more probable for a tank to move
along a road than into a forest.
It is also more probable to move into a position
that has a high military significance, e.g.,
with own units nearby, on top of a hill, etc. 
The probabilities
are also modified by smell/pheromone traces
left there by other ants.
Initially, there is no smell anywhere in the map. As soon
as an ant reaches a favorable military position (e.g., one
of the target
units), it stops and distributes a smell along the way it
took to reach its goal. 

An ant can represent any of several different types
of enemy units, from an infantry-squad to a
battalion of tanks.
The ants basically perform interacting random walks with
probabilities that are site-dependent. 

An ant at position $\bvec{x}$ uses three kinds of information
to determine its future position. First, we have 
geographical information $T(\bvec{y},\bvec{x})$
that simply says how long
it would take the ant to reach each of the neighboring
sites of $\bvec{x}$. This information is predetermined and
comes from a terrain database of the battlefield. It is different
for ants that represent different kinds of enemy units.
This information is taken as fixed in our current simulations,
but it is straightforward to change this in real-time
in order to implement changes in accessibility due to
war activity (e.g., a bombed bridge should be reflected
in this information).

\begin{figure}
\small
\begin{boxedminipage}{\columnwidth}
\begin{enumerate}
\item while maximum time not reached
\begin{enumerate}
\item for all ants $i$
\begin{enumerate}
\item set $\bvec{x}$=current position of ant $i$
\item randomly select a neighbor of $\bvec{x}$ using equation~\ref{walkeq}
and move ant $i$ there
\item if ant $i$ at target then
\begin{enumerate}
\item update smell for all sites visited by ant $i$
according to equation~\ref{smelleq}
\item kill ant $i$
\end{enumerate}
\end{enumerate}
\item if $S(\bvec{x})$ has not changed, exit loop
\end{enumerate}
\item output smell as effective potential
\end{enumerate}
\end{boxedminipage}
\caption{Pseudo-code for the ANTS algorithm.}
\label{pseudo}
\end{figure}

The second component, denoted
$F(\bvec{y})$, is related to the strategic importance
of different locations $\bvec{y}$. This is highest
where the targets that the ants try to reach (i.e., our
units) are. A human operator could change this field
using their intuition and experience of where the
important part of the battle will take place.

The third part is the smell distributed by other ants,
$S(\bvec{y},t)$. At $t=0$, this is initialized to 0 for all $\bvec{y}$:
\begin{equation}
S(\bvec{y},0) = 0  .
\end{equation}
The smell $S$ will be updated during the run of the algorithm, see
equation~\ref{smelleq} below.

The probability to go to a site $\bvec{y}$ from $\bvec{x}$ at time $t$
is thus given by
\begin{equation}
p(\bvec{y},\bvec{x},t) = 0
\end{equation}
if $\bvec{x}$ and $\bvec{y}$ are not nearest neighbors, and
\begin{equation}
p(\bvec{y},\bvec{x},t) \propto 
\frac{1}{T(\bvec{y},\bvec{x})}
+ \omega_s S(\bvec{y},t) + \omega_f F(\bvec{y})
\label{walkeq}
\end{equation}
otherwise. In equation~\ref{walkeq},
$T(\bvec{y},\bvec{x})$ is the geographical information regarding
the time needed for the ant to move from $\bvec{x}$ to $\bvec{y}$,
$S$ and $F$ are the smell and value fields introduced above,
and $\omega_s=1$ and $\omega_f=1$ are weights determining the
relative importances of the different fields. The constant of
proportionality is determined by requiring that
summing over all $\bvec{y}$ gives unit probability.

A possible addition to the $F$ field is to include
also information on visibility and
range of fire at different points, as is done 
in~\cite{richbourg}. The drawback of doing this is that it adds
to the storage and/or computation requirements
for the background fields, thus destroying the
attractive simplicity of the ants model.

In order to avoid loops, we added a small bias against
moving back to where the ant came from. If the new position
is equal to the ant's old position, a new random
number is drawn and a new position is determined from it.
This reduces the probability for the ant to go back in its
own track quadratically. We choose not to completely disallow
back-moves in order to avoid an ant getting
trapped at a location with only one viable exit.
The random number generator used in all simulations
was Matlab's {\bf rand}-function.

All ants are time-evolved in parallel. As soon as an ant
reaches a target site it will stop and a trace
of smell will be distributed along the way it took to reach this
target. If the path that the ant has traveled is given
by the sequence $\bvec{x}_i$ for $i=0, \ldots, M$
(with $\bvec{x}_0$ equal to the starting position), we change the
smell field according to
\begin{equation}
S(\bvec{x}_j,t+1) = S(\bvec{x}_j,t) + \frac{j}{M} .
\label{smelleq}
\end{equation}
For sites $\bvec{x}$ not visited by the ant, the old value for $S$
is propagated to time $t+1$.
This smell will be used by other ants to determine
their future time-evolution (and also displayed to the user
as an indication of where the interesting areas in the map are).
A number of possible extensions can be made here: the ant could
continue from the target when it has distributed its smell or it
could be restarted at the start position.

Pseudo-code for the
algorithm is shown in figure~\ref{pseudo}. 

The output of the program is not the final positions of the ants
but rather the effective potential $S(\bvec{x})$ determined by the
distribution of smell on the map. This distribution
will of course change as more and more ants reach the targets. This
is an important feature of the algorithm: it will run in real-time
and provide incrementally better and better approximations
to the threat analysis.

\section{Results}
\label{gottrora}

\begin{figure}[t]
\vspace{2.5mm}
\centerline{\includegraphics[width=.75 \columnwidth]{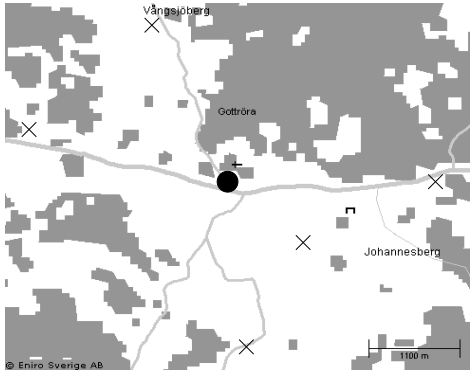}}
\vspace{2mm}
\caption{The example map used for our tests. The black circle
marks the location of the enemy unit whose movement we are trying to
predict, black x:es show locations of own units. The different colors
of the terrain indicate different mobilities for the enemy unit.}
\label{gottrorakarta5}
\end{figure}

The test scenario presented here takes place on the map shown
in figure~\ref{gottrorakarta5}, where we also show
the start position (black circle, close to the center
of the map) and the positions of five targets
(black x:es, one at the middle of each side of
the square and one below and to the right of the center).
The enemy unit at the black circle is assumed to have
different mobilities on the road (light grey), in
the field (white) and forests (dark grey). In
the simulations presented here, the mobility
in the forests is considerably smaller than that in 
the other types of terrain, leading to almost zero
probability of entering such areas.

\begin{figure}[!t]
\centering
\includegraphics[width=.75 \columnwidth]{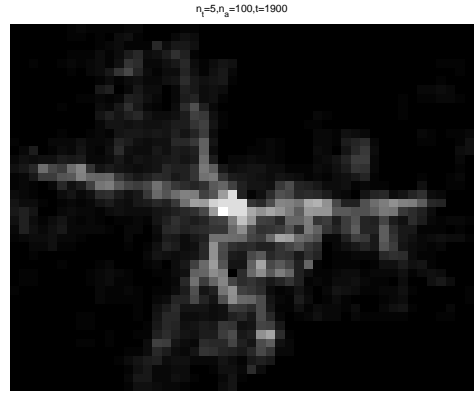}
\vspace{2mm}
\caption{The distribution of smell after 1900 time-steps
for the scenario with 5 targets and 100 ants using a linear gray-scale.
}
\label{origgraymap}
\end{figure}

\begin{figure}[t]
\centering
\includegraphics[width=.75 \columnwidth]{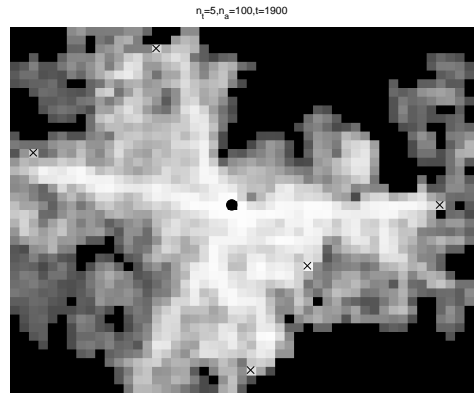}
\vspace{2mm}
\caption{The distribution of smell using a gray-scale
determined by histogram equalization after 1900 time-steps
for the scenario with 5 targets and 100 ants.}
\label{st5a100_1900}
\end{figure}

In figure~\ref{origgraymap}, we present the ``smell'' left by the ants that
is the output of the method. Smells of different strengths are presented
using different gray-scales, where white represents the largest amount
of smell in the figure and black the lowest. 
For visibility
we chose to map the smell onto a non-linear gray-scale using
a histogram equalization method 
(see, e.g.,~\cite{gonzaleswoods}) to determine the appropriate mapping from
smell to gray-scale. 
The results of this transformation
for the distribution shown in figure~\ref{origgraymap} is shown in
figure~\ref{st5a100_1900}. 
(In practice, the transformation works like a
logarithmic gray-scale --- more smell-values are mapped to the
same grayness at large smells than at lower. It works almost like
a filter that filters out all smells higher than some threshold. We
found that this gave the best representation of the distribution
for this medium. )

\begin{figure}[t]
\centering
\includegraphics[width=.75 \columnwidth]{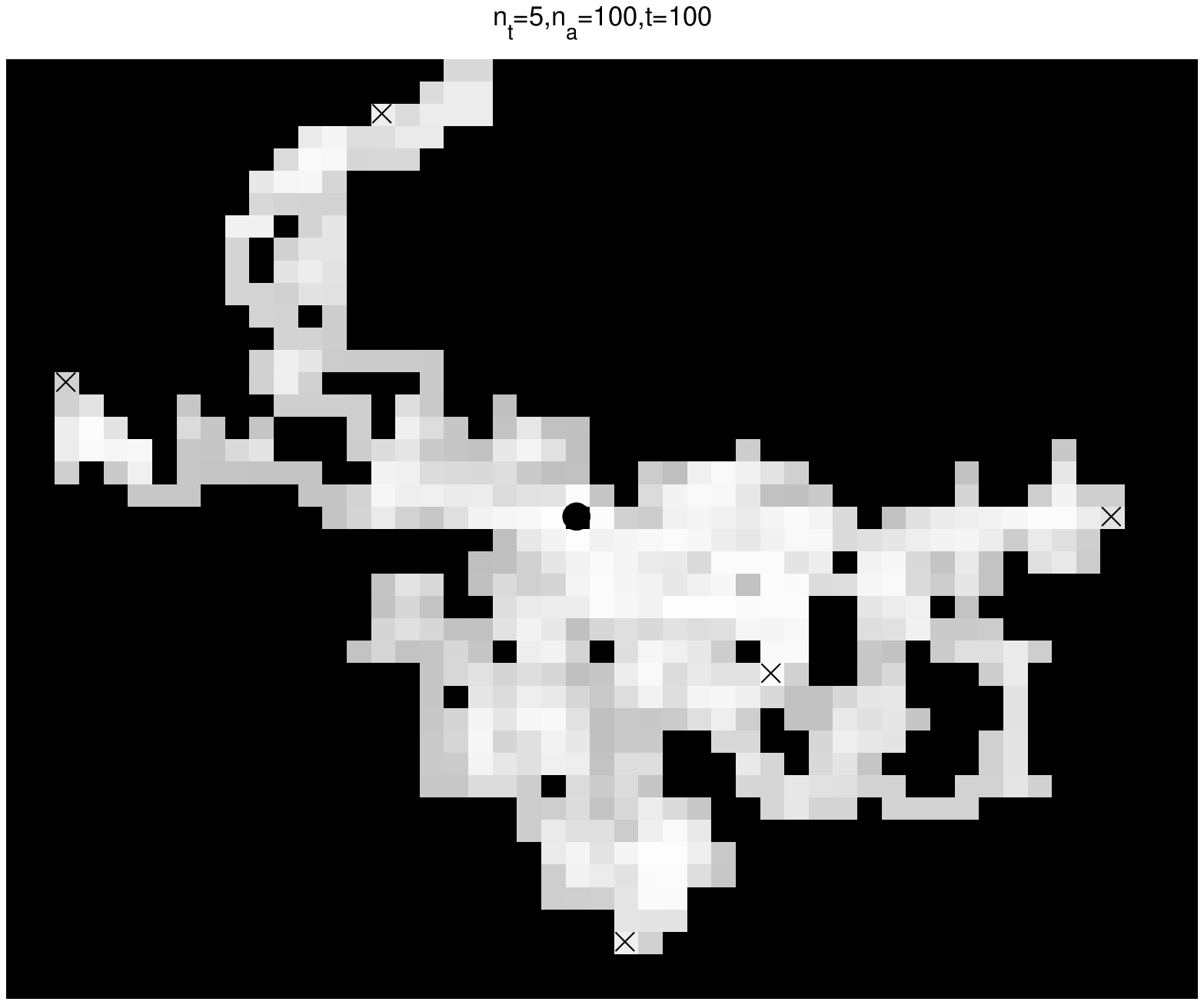}
\vspace{2mm}
\caption{Distribution of smell using a gray-scale
determined by histogram equalization after 100 time-steps
for the scenario with 5 targets and 100 ants.}
\label{st5a100_100}
\end{figure}

\begin{figure}[t]
\centering
\includegraphics[width=.75 \columnwidth]{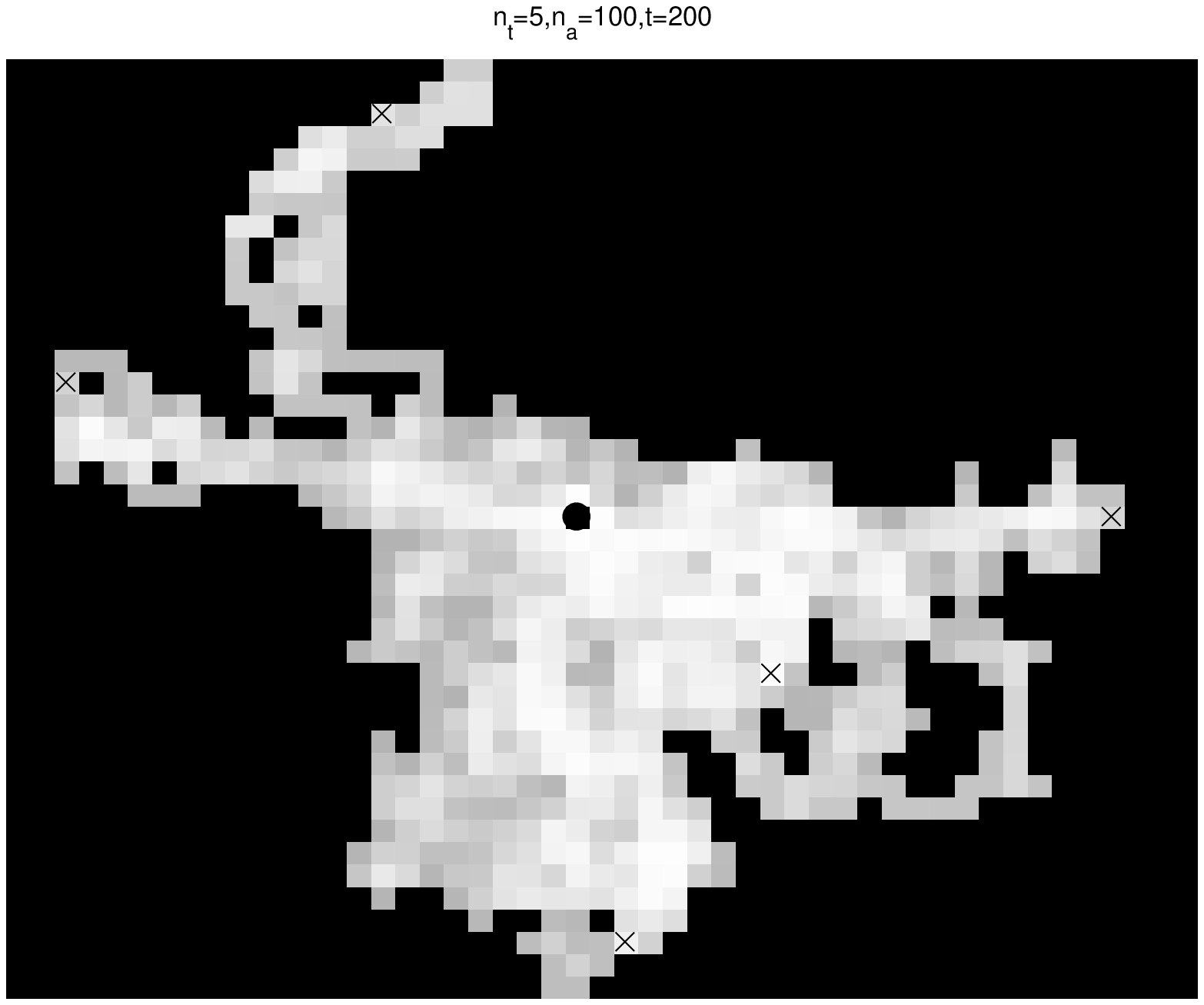}
\vspace{2mm}
\caption{Distribution of smell using a gray-scale
determined by histogram equalization after 200 time-steps
for the scenario with 5 targets and 100 ants.}
\label{st5a100_200}
\end{figure}

\begin{figure}[t]
\centering
\includegraphics[width=.75 \columnwidth]{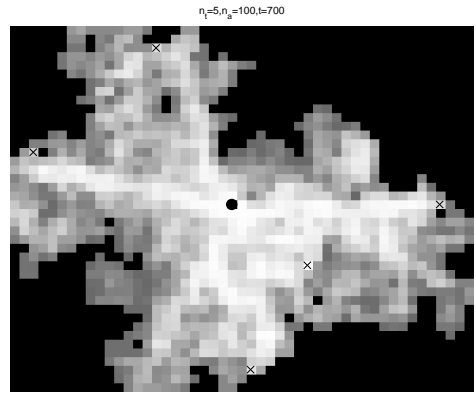}
\vspace{2mm}
\caption{Distribution of smell using histogram equalization after 700 time-steps
for the scenario with 5 targets and 100 ants.}
\label{st5a100_700}
\end{figure}

By comparing figure~\ref{st5a100_1900},
which shows the smell after convergence, to
figures~\ref{st5a100_100} to~\ref{st5a100_700} 
which show the smell
at earlier times, we can see
that the convergence is rather quick for most areas of the map.
Note however the anomalous behavior near the top target.
At times 100 and 200, the distribution of smell indicates that the
enemy would
take a detour going first left along the road and then up through fields
to reach the top. At time
700, the method has discovered the ``correct'' 
avenue of approach to this target.
This is a clear indication of the heuristic nature of the algorithm
showing both the need to run ANTS until convergence and
how it incrementally improves its output.

\begin{figure}[t]
\centering
\includegraphics[width=.75 \columnwidth]{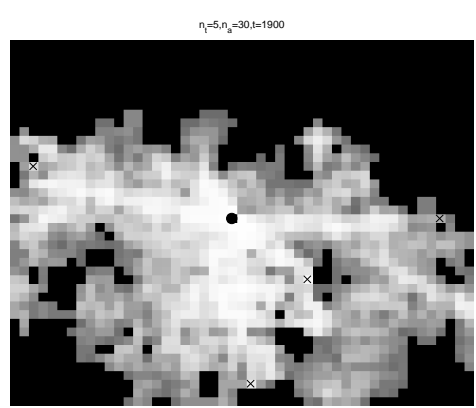}
\vspace{2mm}
\caption{Distribution of smell using a gray-scale determined by histogram equalization after 1900 time-steps
for the scenario with 5 targets and 30 ants.}
\label{st5a30_1900}
\end{figure}

We have also run tests using different numbers of ants. 
In figure~\ref{st5a30_1900}, we show the same
scenario but using just 30 ants. It is clear that this is too small
a number of ants to be able to provide an accurate risk analysis:
the upper target can not be found since the ants get stuck
near the center.

\begin{figure}
\centering
\includegraphics[width=.75 \columnwidth]{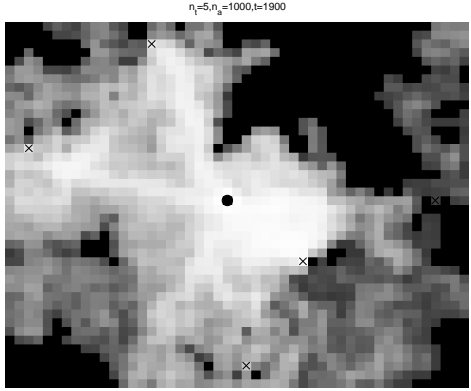}
\vspace{2mm}
\caption{Distribution of smell using a gray-scale determined by histogram equalization after 1900 time-steps
for the scenario with 5 targets and 1000 ants.}
\label{st5a1000_1900}
\end{figure}

Figure~\ref{st5a1000_1900} shows that using
too many ants, in this case $10^3$, also does not lead to a
good convergence of the smell. In this case it is probably
due to too much smell being released near the start, which causes
many of the ants to get trapped here. 
After some experimentation, we tentatively
recommend using on the order of 100 ants for each
simulation. If there are many targets, it is better to divide the targets
into sets containing on the order of $5$ targets each and run a separate
simulation for each set. The results can then be combined for display
in the command and control system.

We have also made tests varying the number and locations of the targets
and found similar results to those in the scene presented here.
As the number of targets and start positions increases,
the probability distribution for future locations will be
fuzzier and fuzzier. This also means that it will we harder and harder for a human
analyst to determine what is happening. The ANTS algorithm provides hints for where the human
should concentrate their attention. 
For analyzing a complex scene with perhaps dozens or hundreds of interesting goals,
it is best to use only a few targets (2-5) at a time and instead run
several simulations and combine their output. In this way,
the ANTS method will be able
to provide a rough guide to where the human analyst should focus their attention.

None of the simulations presented here took more
than a minute to run using Matlab 6.5 on an 1.4GHz AMD Athlon CPU,
and most of the CPU-time was spent in Matlab's functions for
displaying graphics .

\section{Visualizing possible avenues of approach}
\label{presentation}

\begin{figure}[t]
\centering
\leavevmode
\includegraphics[width=.75 \columnwidth]{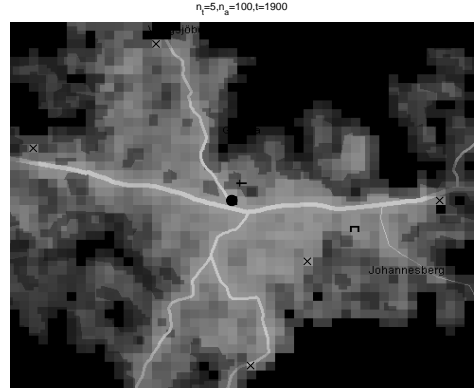}
\vspace{2mm}
\caption{One way of combining histogram-equalized smell and map
reminiscent of representations used in computer games. Data is
shown
for the scenario with 5 targets and 100 ants
after 1900 time-steps.}
\label{t5a100_1900}
\end{figure}

In order to show avenues of approach
on the map, we converted the map to a gray-scale representation,
using darker shades of gray for areas with smaller mobilities.
All subsequent figures show what this map looks like using
the smell as a filter to change the gray-scale. We used
two different ways of combining the map and the smell.
In the first (shown in figure~\ref{t5a100_1900}), the grayness
of a pixel is the product of the map's grayness at that
location and the grayness determined by the
smell. In practice, this means that
areas of the map where no ant has left any smell will be
blacked-out, while those areas that have the most smell
(i.e, that the ANTS algorithm consider most interesting) will
appear normal. The purpose of this is to draw the user's
attention to the avenues of approach,
while the black portions require
less attention.
It is instructive to compare this
with the way the map in computer strategy games
like {\em Civilization} starts out
black and then becomes visible only after the area has
been explored by the player.

\begin{figure}[t]
\centering
\leavevmode
\includegraphics[width=.75 \columnwidth]{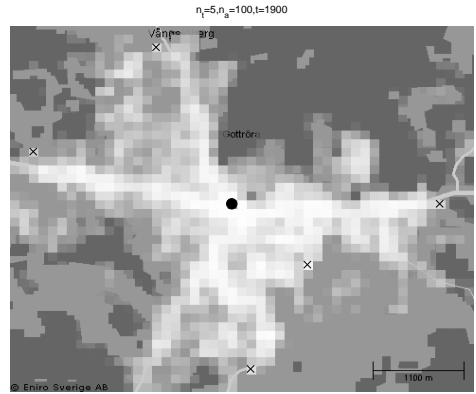}
\vspace{2mm}
\caption{This figure shows max of histogram-equalized smell and map after 1900 time-steps
for the scenario with 5 targets and 100 ants.}
\label{mt5a100_1900}
\end{figure}

The second way of combination (figure~\ref{mt5a100_1900}) 
simply displays the maximum
of the map-value and the smell-value at each pixel. This is a better
representation since it shows more clearly where the ants move and
is less influenced by the roads. The combination using
multiplication is better for showing what areas the enemy
can reach, while the max-combination better shows the relative
differences in occupation probability, and can hence be used for
determining where to increase surveillance or attack.

Note that the particular way we present the results is of course
not a part of the algorithm. 
A real implementation would use colors and shading to combine these
representations and also allow the user to display, e.g.,
only those places where the smell is larger than some
cut-off value.

\section{An ``exact'' potential instead of ANTS}
\label{potentialsektion}

An alternative to using ANTS to determine the effective
potential induced by the targets on the terrain is to
place sources at the target locations and calculate
the exact potential at all locations in the map, taking
into account also the terrain. Assuming Gau\ss ian sources
of strengths $K_n$ and with widths $\sigma_n$, the potential
at location $\bvec{x}$ would then be
\begin{equation}
U_0(\bvec{x}) = \{ \sum_n K_n \exp(\frac{-\parallel \bvec{r}_n - \bvec{x}\parallel^2}{2\sigma_n}) \}
(1-\frac{\parallel \bvec{x}-\bvec{r}_0 \parallel^2}{d^2})  ,
\label{potentialeq}
\end{equation}
where $\bvec{r}_n$ are the goal positions, $\bvec{r}_0$ the start position
and $d$ the diameter of the map.
Note that
this potential as well as all other fields used in this
paper is assumed to live on
the set of points $\bvec{x}$ of a discretized lattice
representation of the area of operations. Note that the exact
expression for the potential in equation~\ref{potentialeq}
is of course completely arbitrary --- we choose this
modified Gau\ss ian for simplicity. It
has nothing to do with the ANTS methods presented here.

Calculating any such potential takes time
\begin{equation}
T_d \sim N_t L^2  ,
\end{equation}
where $N_t$ is the number of targets and $L$ the linear size
of the map. The time needed for the ANTS algorithm to
determine the effective potential is
\begin{equation}
T_a \sim N_a T_c ,
\label{evolutioneq}
\end{equation}
where $N_a$ is the number of ants used and $T_c$ is
the time needed to reach convergence. For diffusion-based
methods without interaction, $T_c$ would scale as
\begin{equation}
T_c \sim \langle \parallel \bvec{r}_t - \bvec{r}_s \parallel^2 \rangle_{\bvec{r}_s} ,
\end{equation}
where $\bvec{r}_s$ is the starting position and the average is
over all target positions $\bvec{r}_t$. The presence of
the smell in equation~\ref{evolutioneq}, however, leads to a
behavior more like that of
super-diffusive dynamics,
\begin{equation}
T_c \sim \max_{\bvec{r}_s} \{ \parallel \bvec{r}_t - \bvec{r}_s \parallel   ^{\alpha} \},
\end{equation}
with $\alpha$ close to 1.
Since 
\begin{equation}
\max_{\bvec{r}_s} \{ \parallel \bvec{r}_t - \bvec{r}_s \parallel^{\alpha} \} \sim L^{\alpha}  ,
\end{equation}
it is clear that ANTS gives a large speed-up over calculating the exact potential.

\begin{figure}[!t]
\centering
\includegraphics[width=.75 \columnwidth]{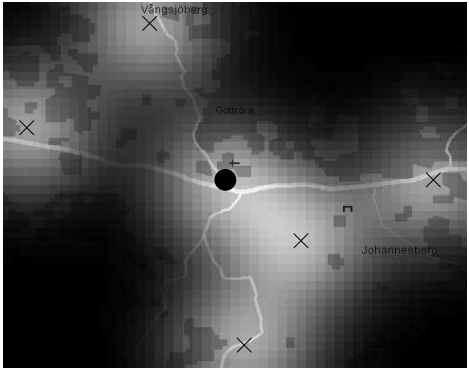}
\vspace{2mm}
\caption{Terrain map modified by potential from targets.}
\label{targetpotential}
\end{figure}

Figure~\ref{targetpotential} shows the terrain map modified
by sources at the target locations, equation~\ref{potentialeq},
using sources of equal strength 4 and with standard deviation 30.
Comparing figure~\ref{targetpotential} to figure~\ref{origgraymap},
it is clear that the effective potential determined by the ANTS method
is a good approximation to this $U_0$. To get a measure of
the speed of convergence of ANTS, we calculated the
discrete Kullback-Leibler~\cite{infotheory} distance
\begin{equation}
K[f,g;t] = \sum_{\bvec{x}} g(\bvec{x},t) \log{\frac{g(\bvec{x},t)}{f(\bvec{x})}} ,
\end{equation}
using normalized distributions
\begin{equation}
f(\bvec{x}) = \frac{U_0(\bvec{x})}{\sum_{\bvec{y}} U_0(\bvec{y}) }
\end{equation}
and
\begin{equation}
g(\bvec{x},t) = \frac{S(\bvec{x},t)}{\sum_{\bvec{y}} S(\bvec{y},t) } .
\end{equation}
We found that the Kullback distance decreased exponentially with time,
stabilizing after about 500 time-steps at a value about an order
of magnitude smaller than at $t=0$. 
Information such as that
displayed in figure~\ref{targetpotential} could be used to
help humans focus on the most important areas of the map.
The exact appearance of equation~\ref{potentialeq}
is of course completely arbitrary. In addition to being much faster,
the ANTS method also does not require us to postulate any
such expression for the potential: it only requires the locations
of targets and the parameters governing the distribution of smell
in equation~\ref{smelleq}.

\section{Discussion}
\label{discuss}

Conventional threat analysis takes the current situation
and uses our knowledge of where our important assets
are to try to predict where the enemy is headed. 
ANTS, in contrast, flips the sides: we try to predict
what the enemy will do by putting ourselves in their position
and determining what we would do, without assuming
that the enemy has global information about our assets
and positions. We argue that
using a local-search method in this way
gives a more robust threat prediction, since
the output is determined by simulating the enemy, not by
trying to guess their objectives.
The ANTS algorithm can be easily adapted
to new information regarding enemy
behavior (by changing equation~\ref{walkeq}).
This is important,
since potential enemies will also have
computer systems to aid them, probably leading to more
surprising tactics.

Another goal of
using ANTS is to minimize the amount of work needed by
humans. Given a terrain map and the locations of enemy forces,
humans can often determine possible
avenues of approach visually,
and then decide where to
concentrate own forces and sensors in defense.
Our method does not aim to completely replace
such human analysis, but can act as a first step by
suggesting such avenues to the analyst.
The output of our program, together with
the output from the situation assessment routines that
are used as inputs to our program, help the human
operators to
focus only on the most important parts of the map.
In addition, since the ANTS algorithm is meant to run
interactively and provide incrementally better distributions
as time goes, it can also be adapted to
give an answer to the question
of what would happen if the enemy suddenly receives some
new information.

\section{Future extensions}
\label{future}

Our current ANTS method can be extended in a number of ways.
It is, for instance, possible to add some movement of
the targets (i.e., our units) or to include
changes in mobility caused by blowing up bridges.

In the simulations presented here, we have used just one
type of object to track at all times. It is straightforward
to extend the method so that it can handle situations
where it is given several different types of
objects (e.g., a platoon of tanks at position $\bvec{x}$
and a company of infantry at position $\bvec{y}$)
as input to get the combined
threat posed by all of these. Ants with different
mobilities should then be started at each of the enemy
positions, and the output should be changed so that it
gives smells for all types of objects. Ants should
here be attracted primarily to smell of its own type,
but also to that of other types. This makes it possible
to model things like tanks following scout patrols
of infantry, or infantry following tanks.

The ANTS method as presented here can also be used for the more
interesting problem of ``opportunity analysis'', i.e.,
to determine what possibilities own forces have given
an accurate situation picture.
We are planning to study these and other extensions
to ANTS in future work.

\section{Conclusions}
\label{conclusions}

In conclusion, we showed how ANTS can be used to
get a quick worst-case risk assessment.
By simulating the enemy instead
of relying on static assumptions of their objectives,
we obtain a method that is more robust if
the enemy also uses computer-assisted command and control
systems.
The ANTS method should be integrated in a command and control
system and provide a first, real-time indication
of avenues of approach.
More thorough analysis methods should also
be part of this system and will give an updated
more exact picture at some later time. The system
should also contain modules that automatically compare
the quick picture with the reliable one, and warns
the human operators when they differ by too much.

The possible avenues of approach that is the output
of the ANTS method can also aid
in sensor allocation and management, to help
determine which areas should
be surveyed by sensors such as UAV's.

{\small {\bf Acknowledgements: } We thank Christian M\aa rtenson
and Johan Schubert for helpful discussions.}


\end{document}